# THERMODYNAMICS OF A BROWNIAN BRIDGE POLYMER MODEL IN A RANDOM ENVIRONMENT[1]

Servet MARTÍNEZ[1] and Dimitri PETRITIS[2]

1. Universidad de Chile, Facultad de Ciencias Físicas y Matemáticas,
Departamento de Ingeniería Matemática, Casilla 170-3 Correo 3, Santiago, Chile.
e-mail: smartine@uchcecvm.cec.uchile.cl

2. Institut de Recherche Mathématique, Université de Rennes I and CNRS URA 305,
Campus de Beaulieu, 35042 Rennes Cedex, France.
e-mail: petritis@levy.univ-rennes1.fr

9 March 1995

## Abstract

We consider a directed random walk making either 0 or +1 moves and a Brownian bridge, independent of the walk, conditionned to arrive at point $b$ on time $T$. The Hamiltonian is defined as the sum of the square of increments of the bridge between the moments of jump of the random walk and interpreted as an energy function over the bridge configuration; the random walk acts as the random environment. The thermodynamic limit of the specific free energy is shown to exist and to be self-averaging, *i.e.* it is equal to a trivial — explicitely computed — random variable. An estimate of the asymptotic behaviour of the ground state energy is also obtained.

## 1 Definition of the model and main results

Disordered systems are thoroughly studied nowadays as condensed matter models in the presence of impurities. Interesting questions, both mathematically and physically, concern the sample dependence or independence of intensive quantities like the specific free energy in statistical mechanics of spin glasses or the integrated density of states in spectral theory

---

[1] *1991 Mathematics Subject Classification*: 82B44, 60G15, 60K15, 60J15

or the critical exponents governing the asymptotic behaviour. Closely related models are random walks in random environments; in this context, interesting questions concern the possible modifications of the asymptotic behaviour critical exponents or the unusual normalisation required for the limit theorems to hold [12].

More recently, the methods developed in the statistical mechanical study of disordered systems have been applied to other systems like long protein or RNA molecules. It is argued in [6] that a protein molecule is very much like a random walk with random charges attached at the vertices of the walk; these charges are interacting through local interactions mimicking Lennard-Jones or hydrogen-bond potentials. On the basis of replica trick heuristics, it is claimed in [6] that the protein undergoes a phase transition between an unfolded state and a folded biologically active state.

The purpose of this paper is to rigorously study a model arising in protein conformation. In [9], Kantor and Kardar introduced a model defined on the product space, $\mathcal{W}_N \times \mathcal{Q}_N$, of $N$-steps one-dimensional nearest neighbours random walks

$$\mathcal{W}_N = \{r : \{1, \cdots, N\} \to \mathbb{Z} \text{ with } r_{i+1} - r_i \in \{-1, 1\} \text{ for } 1 \leq i \leq N - 1\},$$

equipped with the uniform probability, with the probability space $(\mathcal{Q}_N, \mathcal{F}, \mathbb{P})$ where a collection $q = (q_i)_{1 \leq i \leq N}$ of symmetric, independent, and identically distributed random variables are defined. For given $(r, q)$, the Hamiltonian used in [9] is given by

$$H_N^{KK}(r, q) = \sum_{1 \leq i < j \leq N} q_i q_j \delta_{r_i, r_j} \qquad (1)$$

and has the interpretation of conformational energy of a protein molecule with $N$ units whose geometrical shape is given by $r$ and whose charges over the constituting units are given by $q$; these charges interact through an ultralocal interaction. In [9], the random variables $q$ are symmetric Bernoulli variables with values in $\{-1, 1\}$. This model is quite degenerate due to the discrete character of the $q$ variables. For that reason, in [4], the model having the same Hamiltonian as that of the formula (1) but with the variables $q$ being independent Gaussian was studied. Both models are however quite intractable; therefore, in [5] a further simplification was proposed: namely, the variables $q$ are independent Gaussian, the Hamiltonian is again given by the same formula (1), but it is defined on the space $\mathcal{D}_N$ of *directed* nearest neighbours random walks

$$\mathcal{D}_N = \{r : \{1, \ldots, N\} \to \mathbb{N}, \text{ with } r_{i+1} - r_i \in \{0, 1\} \text{ for } 1 \leq i \leq N\}$$

instead of being defined on $\mathcal{W}_N$. This latter simplification allowed the use of transfer matrix techniques which, combined with numerical experiments, gave the asymptotic (for large $N$) behaviour of the ground state energy, $\inf_{r \in \mathcal{D}_N} H^{KK}(r, q)$, of this simplified model under the constraint that the total charge, $Q = \sum_{i=1}^{N} q_i$, is conditioned to remain fixed to a constant $b$.

Our aim is to go beyond the ground state energy and to study the thermodynamics of this model. The main difficulty however is to satisfy the constraint $Q = b$, implying a conditioning on this event, known to be a delicate procedure [10]. To avoid such subtleties, we decided to work on the space $\{Q = b\}$. Moreover, from a fundamental point of view, proteins are soft macromolecular structures, being able to vary their geometry although the charges of the constituting units remain fixed by the chemical properties of the molecule. Therefore, it seems more reasonable to consider the charges $q$ as the microscopic configuration and the form (induced by $r$) as the environment. Our starting point is the Hamiltonian (1) considered as a function on the space $\mathcal{Q}_N$ in a random environment $r \in \mathcal{D}_N$. It proved computationally easier to consider the continuous version of the previous model; it is expected however that both discrete and continuous models have the same asymptotic behaviour. We stop here speculative considerations about the physical intrerpretation of our model and define it mathematically.

Let $T$ be a fixed strictly positive integer and $b$ an arbitrary real number. We denote by $\mathcal{C}_{[0,T]}^{0,b}$ the space of continuous real functions $X$ defined on $[0,T]$ such that $X(0) = 0$ and $X(T) = b$. We denote by $(X_t)_{t \in [0,T]}$ the standard Brownian bridge process on the space $\mathcal{C}_{[0,T]}^{0,b}$. Let $(Y_k)_{k \geq 0}$ be a sequence of independent Bernoulli variables, of parameter $p = \mathbb{P}\{Y_1 = 1\} = 1 - \mathbb{P}\{Y_1 = 0\}$, also independent of the Brownian bridge and defined on a probability space $\Omega$. Define, on the probability space $\Omega$, the renewal process, *i.e.* the strictly increasing sequence of positive integers $(U_i)_{i \geq 0}$, with $U_0 = 0$ and recursively $U_k = \inf\{i > U_{k-1} : Y_i = 1\}$. By $N_T = \sup\{i : U_i < T\}$ we denote the last index $i$ such that $U_i < T$. Thus

$$0 = U_0 < U_1 < \cdots < U_{N_T} < T \leq U_{N_T+1} < \cdots$$

We set $U = (U_1, \cdots, U_N)$ and $\Xi = (\Xi_1, \cdots \Xi_N)$ with

$$\Xi_i = X_{U_i} - X_{U_{i-1}}.$$

The energy of our model is given by the Hamiltonian $H_T : \Omega \times \mathcal{C} \to \mathbb{R}^+$, defined by

$$H_T(U, X) = \sum_{i=1}^{N_T} (X_{U_i} - X_{U_{i-1}})^2 + (b - X_{U_N})^2. \tag{2}$$

This Hamiltonian is the continuous analogue of the Hamiltonian (1) considered in [9]. The length of the protein is identified with the parameter $T$ and the Gaussian charges are modelised through the Brownian bridge process. The constraint of fixed total charge $b$ is automatically satisfied by all configurations in $\mathcal{C}_{[0,T]}^{0,b}$. The directed polymer acts as a random environment for the Brownian bridge.

We define now the thermodynamics of the model. Denote by $\mathbb{E}\{\cdot\}$ the mean expected value, by $\mathbb{E}\{\cdot|\mathcal{R}\}$ the mean expected value conditioned to the process or variable $\mathcal{R}$, and by $\mathbb{E}_\mathcal{R}\{\cdot\}$ the mean expected value with respect to the distribution of $\mathcal{R}$. With these notations, the partition function is

$$Z_T(\beta) = \mathbb{E}\{\exp(-\beta H_T)|(N, U)\}. \tag{3}$$

The parameter $T$ plays obviously the *rôle* of volume. The finite volume, specific, quenched free energy is defined, as usual, by

$$f_T(\beta) = -\frac{1}{\beta T} \log Z_T(\beta) \tag{4}$$

and the finite volume, specific, annealed free energy is given by

$$\overline{f}_T(\beta) = -\frac{1}{\beta T} \log \mathbb{E}_{(N,U)} Z_T(\beta). \tag{5}$$

This model, while reminiscent of the system studied in [11], has however a much richer structure since the random environment, given by the renewal process is independent of the bridge.

With the formulation we use, we impose the total charge to be fixed, avoiding thus the usual subtleties [10, 1] of conditionning with respect to the final point. Moreover, the total charge being left as a free parameters from the very beginning, several scaling laws can be tested.

We study the full thermodynamics of this model and we prove the following

**Theorem 1.1** *Denote by $V_i = U_i - U_{i-1}$ the increment of the renewal process. Let $\lim_{T \to \infty} \frac{b^2}{T} = z \in [0, \infty[$. For every inverse temperature $\beta \in \mathbb{R}^+$, for every asymptotic behaviour of the total charge $z \in [0, \infty[$, the quenched specific free energy, $f_T$, converges in distribution, for $T \to \infty$, to a trivial random variable $f_\infty$, where*

$$f_\infty(\beta) = \frac{p}{2\beta} \mathbb{E} \log(1 + 2\beta V_1),$$

*and $p = \mathbb{P}(Y_1 = 1)$.*

**Remark:** The previous theorem is established for a finite limit $z$. It is not difficult to see however that this result remains true even in the case $z = \infty$ provided that the divergence of the ratio $b^2/T$ is not very violent (sublinear).

Besides the thermodynamic behaviour we obtain also results on the scaling of the ground state energy. It is argued, in [9], that when the total charge $b$ scales asymptotically like $b \sim T^x$ (where $f \sim g$ means $\lim_{T \to \infty} \frac{f(T)}{g(T)} = 1$) then the minimum of the energy scales like $\min H \sim T^{\theta(x)}$, where $\theta(x)$ is a critical exponent continuously depending on $x$. These results are based on heuristic arguments and numerical simulations. Here we prove the following

**Theorem 1.2** *Let $0 < p = \mathbb{P}(Y_1 = 0) < 1$, $q = 1 - p$, $b \in \mathbb{R}$, and $T \in \mathbb{N}^+$. Then, we have*

$$-\lim_{\beta \to \infty} \frac{1}{\beta} \log \left[ \mathbb{E} \left( \exp(-\beta H_T) | (N, U) \right) \right] = \frac{b^2}{(N_T + 1)}.$$

**Corollary 1.3** *Under the same conditions as the previous theorem,*

$$\mathbb{E} \left( \inf_{X \in \mathcal{C}_{[0,T]}^{0,b}} H_T(U, X) \right) = \frac{b^2}{pT}(1 - q^T).$$

The previous result suggests that for fixed $p$ and $q = 1 - p$, if $b \sim T^x$, the dominant behaviour of the minimum of the Hamiltonian will behave as $\mathbb{E}[\min_X H] \sim T^{2x}$. For $p$ and $q$ depending on $T$ however, the precise asymptotic behaviour can be changed.

The paper is organised as follows: the next section deals with technical explicit computations and intermediate results on finite volume systems. These results are used in section 3 to obtain the thermodynamic limit of the specific free energy. The zero temperature limit and ground state energy are studied in section 4. In a last section some open problems are presented. The appendices remind some well known results for the paper to be self contained.

## 2 The conditional expectation of the Boltzmann factor

Let $\vec{u} = (u_1, \cdots, u_n)$ be a vector with $0 < u_1 < \cdots < u_n < T$. Denote by $\mathbb{E}(\cdot|n, \vec{u})$ the conditional expectation with respect to the event $\{N = n; U_1 = u_1, \cdots, U_n = u_n\}$ and by $\mu_\Xi = \mathbb{E}(\Xi|n, \vec{u})$ and $Q_\Xi = (\mathbb{E}[\Xi_i\Xi_j - (\mu_\Xi)_i(\mu_\Xi)_j|n, \vec{u}]; i, j = 1, \cdots, n) = \mathrm{Cov}\{\Xi|n, \vec{u}\} = (\mathrm{Cov}\{(\Xi_i, \Xi_j)|\vec{n}, \vec{u}\} : i, j = 1, \cdots, n)$ the mean expected vector and the covariance matrix of $\Xi$ conditioned to the event $\{(N, U) = (n, \vec{u})\}$ respectively. We have

$$\mathbb{E}[\exp(-\beta H_T)|n, \vec{u}] = (2\pi)^{-n/2}|\det Q_\Xi^{-1}|^{1/2} \int_{\mathbb{R}^n} \exp[-\{\beta H_T + \frac{1}{2}(\vec{\xi} - \mu_\Xi)^t Q_\Xi^{-1}(\vec{\xi} - \mu_\Xi)\}]|d\vec{\xi}|.$$

Denote $q = 1 - p$ and let $\phi : \mathbb{N}^+ \times \mathbb{R}^n \to \mathbb{R}$ be an arbitrary integrable function; then we have [7]

$$\mathbb{E}_{(N,U)}(\phi) = \sum_{n=0}^{T-1} p^n q^{T-1-n} \sum_{u_1=1}^{T-1-n} \cdots \sum_{u_i=u_{i-1}+1}^{T-1-(n-i)} \cdots \sum_{u_n=u_{n-1}+1}^{T-1} \mathbb{E}\left(\phi(N, U)|n, \vec{u}\right).$$

Consequently,

$$\mathbb{E}Z_T(\beta) = \sum_{n=0}^{T-1} p^n q^{T-1-n} \sum_{u_1=1}^{T-1-n} \cdots \sum_{u_n=u_{n-1}+1}^{T-1} \mathbb{E}[\exp(-\beta H_T)|n, \vec{u}].$$

**Lemma 2.1** *Set $v_i = u_i - u_{i-1}$ for $i = 1, \cdots, n$ and $\Xi_i = X_{u_i} - X_{u_{i-1}}$ for $i = 1, \cdots, n$. Denote by $\vec{v} = (v_1, \cdots, v_n)$ and for $v_i \neq 0, \forall i$, denote by $\vec{v}^{-1} = (v_1^{-1}, \cdots, v_n-1)$. Let $D(\vec{v})$ the diagonal matrix given by $D(\vec{v})_{ij} = v_i \delta_{ij}$ and, for every $a \in \mathbb{R}$, denote by $\tilde{a}$ the constant matrix whose coefficients $\tilde{a}_{ij} = a$ for all $i$ and $j$. Then, the inverse of the covariance matrix $Q_\Xi$ exists and is given by*

$$Q_\Xi^{-1} = D(\vec{v}^{-1}) + \left(T - \sum_{i=1}^n v_i\right)^{-1} \tilde{1} = D(\vec{v}^{-1}) + (T - u_n)^{-1}\tilde{1}. \tag{6}$$

*Moreover,*

$$\det Q_\Xi^{-1} = \prod_{j=1}^n v_j^{-1} \left(1 + (T - u_n)^{-1} u_n\right) = \left(\prod_{j=1}^n v_j^{-1}\right) T (T - u_n)^{-1}. \tag{7}$$

*Proof:* From the elementary properties of the Brownian bridge reminded in appendix A, we get

$$(\mu_\Xi)_i = \mathbb{E}[\Xi_i | n, \vec{u}] = \frac{b}{T} v_i$$

and

$$\text{Cov}\,[(\Xi_i, \Xi_j) | n, \vec{u}] = v_i \delta_{ij} - \frac{1}{T} v_i v_j \text{ for } i, j = 1, ..., n.$$

Introducing the matrix $S(\vec{v})$ whose coefficients are given by $S(\vec{v})_{ij} = v_i v_j$, we can write the previous results in the matrix form

$$\mu_\Xi = \frac{b}{T} \vec{v} \quad \text{and} \quad Q_\Xi = D(\vec{v}) - \frac{1}{T} S(\vec{v}).$$

Since $\frac{1}{T} \sum_{i=1}^n v_i = \frac{u_n}{T} < 1$, we can use lemma (C.1) to assert that $Q_\Xi$ is invertible and more precisely, obtain the explicit expression of the inverse

$$Q_\Xi^{-1} = D(\vec{v}^{-1}) + \left(T - \sum_{i=1}^n v_i\right)^{-1} \widetilde{1} = D(\vec{v}^{-1}) + (T - u_n)^{-1} \widetilde{1}. \tag{8}$$

Now use of lemma (B.1) allows to explicitely compute the determinant by the formula

$$\det Q_\Xi^{-1} = \prod_{j=1}^n v_j^{-1} \left(1 + (T - u_n)^{-1} u_n\right) = \left(\prod_{j=1}^n v_j^{-1}\right) T (T - u_n)^{-1}.$$

□

**Lemma 2.2** *On the event* $\{N = n; U_1 = u_1, \cdots, U_n = u_n\}$, *the Hamiltonian can be expressed by the formula*

$$H_T(\vec{u}, \Xi) = \Xi^t \Xi + b^2 - 2b\Xi^t \vec{1} + \Xi^t \widetilde{1} \Xi = \Xi^t (I + \widetilde{1})\Xi - 2b\Xi^t \vec{1} + b^2,$$

*where for $a \in \mathbb{R}$ we put $\vec{a} = (a, \cdots, a) \in \mathbb{R}^n$ to denote the constant vector of $\mathbb{R}^n$ whose components are equal to $a$.*

*Proof:* On the event $\{N = n; U_1 = u_1, \cdots, U_n = u_n\}$, the energy is given by $H_T(\vec{u}, X) = \sum_{i=1}^n (X_{u_i} - X_{u_{i-1}})^2 + (b - X_{u_n})^2$. Since $X_{u_n} = \sum_{i=1}^n \Xi_i$, we find

$$H_T(\vec{u}, \Xi) = \sum_{i=1}^n \Xi_i^2 + \left(b - \sum_{i=1}^n \Xi_i\right)^2 = \sum_{i=1}^n \Xi_i^2 + b^2 - 2b \sum_{i=1}^n \Xi_i + \left(\sum_{i=1}^n \Xi_i\right)^2.$$

Using the compact vector notation and denoting by $(\cdot)^t$ the transpose of $(\cdot)$ (matrix or vector), we get the desired formula. □

Let us examine now the quadratic form

$$M(\vec{u}, \vec{\xi}) = \beta H_T(\vec{u}, \vec{\xi}) + \frac{1}{2}(\vec{\xi} - \mu_\Xi)^t Q_\Xi^{-1}(\vec{\xi} - \mu_\Xi).$$

Since $Q_\Xi^{-1}$ is symmetric, we have

$$M(\vec{u}, \vec{\xi}) = \frac{1}{2}\vec{\xi}^t(Q_\Xi^{-1} + 2\beta I + 2\beta\tilde{1})\vec{\xi} - \vec{\xi}^t(2b\beta\vec{1} + Q_\Xi^{-1}\mu_\Xi) + \beta b^2 + \frac{1}{2}\mu_\Xi^t Q_\Xi^{-1}\mu_\Xi.$$

Put $Q(\beta) = Q_\Xi^{-1} + 2\beta I + 2\beta\tilde{1}$; from (8), we get

$$Q(\beta) = D(\vec{v})^{-1} + 2\beta I + \left((T - u_n)^{-1} + 2\beta\right)\tilde{1}.$$

With the compact vector notation introduced so far,

$$Q(\beta) = D(\vec{v}^{-1} + 2\vec{\beta}) + \gamma(\beta)\tilde{1}$$

with

$$\gamma(\beta) = (T - u_n)^{-1} + 2\beta. \tag{9}$$

Assuming that $\beta \geq 0$, the matrix $Q(\beta)$ is positive definite as it is shown in appendix B. Therefore, there exists a matrix $\Gamma(\beta)$ that is the square root of $Q(\beta)$. The notation $\Gamma(\beta)$ is an abridged writing of the more complete one, introduced in appendix B; namely,

$$\Gamma(\beta) \equiv \Gamma(\vec{v}^{-1} + 2\vec{\beta}, \gamma(\beta));$$

we stick however on this simplified notation not to excessively burden the formulae. Now, it is possible to express $M(\vec{\xi})$ in terms of a vector $\vec{r}(\beta)$ and a real $c(\beta)$ such that for any $\xi \in \mathbb{R}^n$,

$$M(\vec{u}, \vec{\xi}) = \frac{1}{2}\left(\Gamma(\beta)\vec{\xi} - \vec{r}(\beta)\right)^t \left(\Gamma(\beta)\vec{\xi} - \vec{r}(\beta)\right) + c(\beta).$$

Both existence and explicit formulae for $\vec{r}(\beta)$ and $c(\beta)$ are obtained by remarking that $\Gamma(\beta)$ is a symmetric matrix and hence, by identification, $\Gamma(\beta)\vec{r}(\beta) = 2b\beta\vec{1} + Q_\Xi^{-1}\mu_\Xi$ which implies

$$\vec{r}(\beta) = \Gamma(\beta)^{-1}(2b\beta\vec{1} + Q_\Xi^{-1}\mu_\Xi) \tag{10}$$

and

$$c(\beta) = \beta b^2 + \frac{1}{2}\mu_\Xi^t Q_\Xi^{-1}\mu_\Xi - \frac{1}{2}\vec{r}(\beta)^t\vec{r}(\beta). \tag{11}$$

**Lemma 2.3** *Assuming $\beta \geq 0$ and using the notation introduced so far, we have*

$$\mathbb{E}\left(\exp(-\beta H_T)|N, U\right) = \psi(N, U),$$

*where*

$$\psi(n, \vec{u}) \equiv \mathbb{E}[\exp(-\beta H_T)|n, \vec{u}]$$

$$= \left\{\prod_{j=1}^{n}\left(\frac{1}{1+2\beta v_j}\right)\right\}^{\frac{1}{2}} \left\{\frac{T(T-u_n)^{-1}}{\left(1+\gamma(\beta)\left(\sum_{j=1}^{n}(v_j^{-1}+2\beta)^{-1}\right)\right)}\right\}^{\frac{1}{2}} \exp(-c(\beta)).$$

*Proof:* Writing explicitely the integrations involved in the conditional expectation, we obtain

$$\mathbb{E}[\exp(-\beta H_T)|n, \vec{u}] = (2\pi)^{-n/2} |\det Q_\Xi^{-1}|^{\frac{1}{2}} e^{-c(\beta)} \int_{\mathbb{R}^n} \exp\left[-\frac{1}{2}\left(\Gamma(\beta)\vec{\xi}-\vec{r}(\beta)\right)^t \left(\Gamma(\beta)\vec{\xi}-\vec{r}(\beta)\right)\right] |d\vec{\xi}|.$$

Make the change of variables $\vec{\xi}' = \Gamma(\beta)\vec{\xi} - \vec{r}(\beta)$; this introduces a Jacobian in the volume element, of the form $|d\vec{\xi}| = |\det \Gamma(\beta)|^{-1}|d\vec{\xi}'|$. Then perform explicitly the Gaussian integration over the Brownian bridge to get

$$\mathbb{E}[\exp(-\beta H)|n, \vec{u}] = |\det Q_\Xi^{-1}|^{\frac{1}{2}} |\det \Gamma(\beta)|^{-1} \exp[-c(\beta)].$$

Now make use the corollary B.2 for the determinant of the matrix $\Gamma(\beta)$, namely

$$|\det \Gamma(\beta)| = \left(\prod_{j=1}^{n}(v_j^{-1}+2\beta)\right)^{\frac{1}{2}} \left(1+\gamma(\beta)\sum_{j=1}^{n}(v_j^{-1}+2\beta)^{-1}\right)^{\frac{1}{2}}.$$

Combination with the formula (7) for the determinant of the matrix $Q_\Xi$ leads to the desired result. $\square$

Next we need an explicit evaluation of the parameter $c(\beta)$. This evaluation is quite straightforward but rather cumbersome; several auxiliary quantities are introduced that allow to express this parameter. Let

$$A_1(\beta) = \frac{2b^2\beta^2}{\left[\gamma(\beta)+\left(\sum_{i=1}^{n}(v_i^{-1}+2\beta)^{-1}\right)^{-1}\right]}$$

$$A_2(\beta) = \frac{2b^2\beta}{T}\left(\sum_{i=1}^{n}(v_i^{-1}+2\beta)^{-1}\right)$$

$$A_3(\beta) = \frac{2b^2\beta}{T}\left(\frac{1}{\gamma(\beta)\left(\sum_{i=1}^{n}(v_i^{-1}+2\beta)^{-1}\right)}+1\right)^{-1}\left\{(T-u_n)^{-1}\gamma(\beta)^{-1}u_n - \sum_{i=1}^{n}(v_i^{-1}+2\beta)^{-1}\right\}$$

$$A_4(\beta) = \frac{b^2}{2(T-u_n)^2}\left(\gamma(\beta)+\left(\sum_{i=1}^{n}(v_i^{-1}+2\beta)^{-1}\right)^{-1}\right)^{-1}.$$

**Lemma 2.4** *With the notations introduced above, and under the conditions of the previous lemma, we have*

$$c(\beta) = \beta b^2 + \frac{1}{2}\frac{b^2}{T}\frac{u_n}{T - u_n} - A_1(\beta) - A_2(\beta) - A_3(\beta) - A_4(\beta).$$

*Proof:* Start from the expression (11) for $c(\beta) = \beta b^2 + \frac{1}{2}\mu_\Xi^t Q_\Xi^{-1}\mu_\Xi - \frac{1}{2}\vec{r}(\beta)^t\vec{r}(\beta)$. First evaluate the term $\frac{1}{2}\mu_\Xi^t Q_\Xi^{-1}\mu_\Xi$; using the expression given in (8) for the inverse of the matrix $Q_\Xi$, we get

$$\frac{1}{2}\mu_\Xi^t Q_\Xi^{-1}\mu_\Xi = \frac{1}{2}\frac{b^2}{T^2}\left(\vec{v}^t D(\vec{v})^{-1}\vec{v} + (T - u_n)^{-1}\vec{v}^t\tilde{1}\vec{v}\right)$$

$$= \frac{1}{2}\frac{b^2}{T}\frac{u_n}{T - u_n}.$$

Compute now the term $\frac{1}{2}\vec{r}(\beta)^t\vec{r}(\beta)$. From (10), the symmetry of the matrices, and the fact that $\Gamma(\beta)^{-2} = Q(\beta)^{-1}$, we get

$$\frac{1}{2}\vec{r}(\beta)^t\vec{r}(\beta) = 2b^2\beta^2\vec{1}^t Q(\beta)^{-1}\vec{1} + 2b\beta\vec{1}^t Q(\beta)^{-1}Q_\Xi^{-1}\mu_\Xi + \frac{1}{2}\mu_\Xi^t Q_\Xi^{-1}Q(\beta)^{-1}Q_\Xi^{-1}\mu_\Xi \quad (12)$$

Recall that $Q(\beta) = D(\vec{v}^{-1} + 2\vec{\beta}) + \gamma(\beta)\tilde{1}$ with $\gamma(\beta)$ given by (9). Since we restrict ourselves to $\beta \geq 0$ we can apply Lemma C.1 to get

$$Q(\beta)^{-1} = D(\vec{w}(\beta)) - h(\beta)S(\vec{w}(\beta)),$$

where

$$\vec{w}(\beta) = (\vec{v}^{-1} + 2\vec{\beta})^{-1}$$

and

$$h(\beta) = \left(\gamma(\beta)^{-1} + \sum_{i=1}^n w_i(\beta)\right)^{-1}.$$

Then the first term of the right hand side of (12) is equal to

$$2b^2\beta^2(\vec{1}^t D(\vec{w}(\beta))\vec{1} - h(\beta)\vec{1}^t S(\vec{w}(\beta))\vec{1}) = 2b^2\beta^2\left(\left(\sum_{i=1}^n w_i(\beta)\right) - h(\beta)\left(\sum_{i=1}^n w_i(\beta)\right)^2\right)$$

$$= 2b^2\beta^2\left(\gamma(\beta) + \left(\sum_{i=1}^n (v_i^{-1} + 2\beta)^{-1}\right)^{-1}\right)^{-1}$$

so that

$$2b^2\beta^2\vec{1}^t Q(\beta)^{-1}\vec{1} = A_1(\beta). \quad (13)$$

For evaluating the second term of (12), we first compute $Q(\beta)^{-1}Q_Z^{-1}$. On that purpose, introduce the following notation: for a vector $\vec{\ell}$ denote by $L(\vec{\ell})$ the matrix

$$L(\vec{\ell}) = \begin{pmatrix} \ell_1 & \cdots & \ell_1 \\ \ell_2 & \cdots & \ell_2 \\ \vdots & & \vdots \\ \ell_n & \cdots & \ell_n \end{pmatrix}.$$

Observe that $D(\vec{\ell})\tilde{1} = L(\vec{\ell})$ and $S(\vec{\ell})\tilde{1} = \left(\sum_{i=1}^{n} \ell_i\right) L(\vec{\ell})$. Denote $\vec{w}\vec{v} = (w_1v_1, ..., w_nv_n)$. Then we get,

$$\begin{aligned}
Q(\beta)^{-1}Q_\Xi^{-1} &= \{D(\vec{w}(\beta)) - h(\beta)S(\vec{w}(\beta))\}\left\{D(\vec{v}^{-1}) + (T-u_n)^{-1}\tilde{1}\right\} \\
&= D(\vec{w}(\beta))D(\vec{v}^{-1}) + (T-u_n)^{-1}\left(1 - h(\beta)\left(\sum_{i=1}^{n} u_i(\beta)\right)\right)L(\vec{w}(\beta)) \\
&\quad - h(\beta)S(\vec{w}(\beta))D(\vec{v}^{-1}).
\end{aligned}$$

Since $1 - h(\beta)\left(\sum_{i=1}^{n} u_i(\beta)\right) = \gamma(\beta)^{-1}\left(\gamma(\beta)^{-1} + \left(\sum_{i=1}^{n} u_i(\beta)\right)\right)^{-1}$ we get that

$$\begin{aligned}
Q(\beta)^{-1}Q_\Xi^{-1} &= D((\vec{v}^{-1} + 2\vec{\beta})^{-1}\vec{v}^{-1}) + \frac{L((\vec{v}^{-1} + 2\vec{\beta})^{-1})}{(T-u_n)\gamma(\beta)\left(\gamma(\beta)^{-1} + \sum_{i=1}^{n}(v_i^{-1} + 2\beta)^{-1}\right)} \\
&\quad - \frac{S((\vec{v}^{-1} + 2\vec{\beta})^{-1})D(\vec{v}^{-1})}{\left(\gamma(\beta)^{-1} + \sum_{i=1}^{n}(v_i^{-1} + 2\beta)^{-1}\right)}.
\end{aligned}$$

The second term of (12) now reads

$$\begin{aligned}
2b\beta\vec{1}^t Q(\beta)^{-1}Q_\Xi^{-1}\mu_\Xi &= \frac{2b^2\beta}{T}(\vec{1}^t Q(\beta)^{-1}Q_\Xi^{-1}\vec{v}) \\
&= \frac{2b^2\beta}{T}\left\{\vec{1}^t D(\vec{w}(\beta))D(\vec{v}^{-1})\vec{v} + \frac{\vec{1}^t L(\vec{w}(\beta))\vec{v}}{(T-u_n)\gamma(\beta)\left(\gamma(\beta)^{-1} + \sum_{i=1}^{n} w_i(\beta)\right)}\right. \\
&\quad \left. - \frac{\vec{1}^t S(\vec{w}(\beta))D(\vec{v}^{-1})\vec{v}}{\left(\gamma(\beta)^{-1} + \sum_{i=1}^{n} w_i(\beta)\right)}\right\}.
\end{aligned}$$

We have the following identities

$$\vec{1}^t D(\vec{w}(\beta))D(\vec{v}^{-1})\vec{v} = \vec{1}^t D(\vec{w}(\beta))\vec{1} = \sum_{i=1}^{n} w_i(\beta),$$

$$\vec{1}^t L(\vec{w}(\beta))\vec{v} = \left(\sum_{i=1}^{n} w_i(\beta)\right)\vec{1}^t\vec{v} = \left(\sum_{i=1}^{n} w_i(\beta)\right)u_n,$$

and

$$\vec{1}^t S(\vec{w}(\beta))D(\vec{v}^{-1})\vec{v} = \vec{1}^t S(\vec{w}(\beta))\vec{1} = \left(\sum_{i=1}^{n} w_i(\beta)\right)^2.$$

Combining, we obtain,

$$2b\beta\vec{1}^t Q(\beta)^{-1}Q_\Xi^{-1}\mu_\Xi = \frac{2b^2\beta}{T}\left(\sum_{i=1}^{n} w_i(\beta)\right) + \frac{2b^2\beta}{T}\left(\sum_{i=1}^{n} w_i(\beta)\right)\frac{(T-u_n)^{-1}\gamma(\beta)^{-1}u_n - \sum_{i=1}^{n} w_i(\beta)}{\gamma(\beta)^{-1} + \sum_{i=1}^{n} w_i(\beta)}$$

so that, finally, the second term of (12) reads

$$2b\beta \vec{1}^t Q(\beta)^{-1} Q_\Xi^{-1} \mu_\Xi = A_2(\beta) + A_3(\beta). \tag{14}$$

It remains to evaluate the last term of (12); rewrite,

$$\frac{1}{2}\mu_\Xi^t Q_\Xi^{-1} Q(\beta)^{-1} Q_\Xi^{-1} \mu_\Xi = \frac{1}{2}\frac{b^2}{T^2} \vec{v}^t Q_\Xi^{-1} Q(\beta)^{-1} Q_\Xi^{-1} \vec{v}$$

and

$$\begin{aligned} \vec{v}^t Q_\Xi^{-1} &= \vec{v}^t \left( D(\vec{v})^{-1} + (T - u_n)^{-1} \widetilde{1} \right) \\ &= \vec{1}^t + (T - u_n)^{-1} u_n \vec{1}^t \\ &= (T - u_n)^{-1} \vec{1}^t. \end{aligned}$$

From this equation and its symmetric couterpart,

$$\begin{aligned} \frac{1}{2}\mu_\Xi^t Q_\Xi^{-1} Q(\beta)^{-1} Q_\Xi^{-1} \mu_\Xi &= \frac{1}{2}b^2 (T - u_n)^{-2} \vec{1}^t Q(\beta)^{-1} \vec{1} \\ &= \frac{1}{2}b^2 (T - u_n)^{-2} \left\{ \vec{1}^t D(\vec{w}(\beta)) \vec{1} - h(\beta) \vec{1}^t S(\vec{w}(\beta)) \vec{1} \right\} \\ &= \frac{1}{2}b^2 (T - u_n)^{-2} \left\{ \sum_{i=1}^n w_i(\beta) - h(\beta) \left( \sum_{i=1}^n w_i(\beta) \right)^2 \right\} \\ &= \frac{1}{2}b^2 (T - u_n)^{-2} \left( \gamma(\beta) + \left( \sum_{i=1}^n (v_i^{-1} + 2\beta)^{-1} \right)^{-1} \right)^{-1}, \end{aligned}$$

so that finally

$$\frac{1}{2}\mu_\Xi^t Q_\Xi^{-1} Q(\beta)^{-1} Q_\Xi^{-1} \mu_\Xi = A_4(\beta). \tag{15}$$

$\square$

**Remark:** Since explicit formulae for the conditional expectation with respect to the event

$$\{(N, U) = (n, \vec{u})\}$$

are obtained, the conditional expectation $\mathbb{E}(\cdot | N, U)$ is also explicitly known. In fact, if for an integrable function $F$,

$$\mathbb{E}(F | n, \vec{u}) = \phi(n, \vec{u})$$

then

$$\mathbb{E}(F | N, U) = \phi(N, U).$$

So that, if we replace in the above formulae $n$ with $N_T$ and $u_i$ or $v_i$ with $U_i$ and $V_i$ respecitvely, we obtain the corresponding expressions for the conditional expectations $\mathbb{E}(\cdot | N, U)$.

# 3 Thermodynamic limit of the specific free energy

In this section the $T \to \infty$ limit of the specific free energy will be evaluated. Some notation and well known results on the renewal process are necessarry. Since $U_0 = 0$ and $U_n = \inf\{k > U_{n-1} : Y_k = 1\}$, the variables $Y_i$ being independent and identically distributed, the renewal process $(U_i)$ can be written as a sum of independent geometric variabless $U_n = \sum_{i=1}^{n} V_i$, where $V_i = U_i - U_{i-1}$. For every integer $x \geq 1$, we have that $\mathbb{P}(V_1 = x) = pq^{x-1}$, where $p = \mathbb{P}(Y_1 = 1) = 1 - q$. Therefore, $\mu \equiv \mathbb{E} V_i = \frac{1}{p}$. Of special interest is also the random variable $N_T = \sup\{k : U_k < T\}$. Obviously, $\lim_{T \to \infty} N_T = \infty$. Notice also that although $N_T$ is not a stopping time with respect to the $\sigma$-algebra $\sigma(Y_1, \cdots, Y_T)$, the random variable $N_T + 1$ is, on the contrary, a stopping time.

**Lemma 3.1** *Almost surely, we have,*

$$\lim_{T \to \infty} \frac{N_T}{T} = \frac{1}{\mu} = p.$$

*Proof:* The proof is quite standard (see [2] for instance) and relies on the strong law of large numbers. □

**Lemma 3.2** *The finite volume specific free energy, $f_T$, is a random variable reading*

$$\begin{aligned}
f_T(\beta) &= \frac{1}{2\beta T} \sum_{i=1}^{N_T} \log(1 + 2\beta V_i) + \frac{1}{2\beta T} \log T - \frac{1}{2\beta T} \log(T - U_{N_T}) \\
&\quad - \frac{1}{2\beta T} \log\left[1 + \left((T - U_{N_T})^{-1} + 2\beta\right) \sum_{i=1}^{N_T} (V_i^{-1} + 2\beta)^{-1}\right] + \frac{b^2}{T} \\
&\quad + \frac{1}{2\beta} \frac{U_{N_T}}{T(T - U_{N_T})} - \frac{A(\beta)}{\beta T},
\end{aligned}$$

*where $A(\beta)$ is a random variable such that*

$$\mathbb{E}(A(\beta)|n, \vec{u}) = A_1(\beta) + A_2(\beta) + A_3(\beta) + A_4(\beta)$$

*and $A_1, \ldots, A_4$ are the expressions given in the previous section.*

*Proof:* Recalling the remark on conditional expectations given at the end of the previous section and lemmata (2.3) and (2.4), the present result follows immediately. □

Studying the thermodynamic limit of $f_T$ is equivalent in studying the $T \to \infty$ limit in the above formula. Notice however that besides random terms whose limit must be computed, there are also deterministic terms of the form $b^2/T$ where $b$ is the total charge of the protein, supposed to scale with the size $T$ of the molecule like $b \sim T^\alpha$. It follows that

$$z \equiv \lim_{T \to \infty} \frac{b^2}{T} = \begin{cases} 1 & \text{if } \alpha = 1/2, \\ 0 & \text{if } \alpha < 1/2, \\ \infty & \text{if } \alpha > 1/2. \end{cases}$$

We do not fix the exact behaviour here. We only assume that $z \in [0, \infty]$ in the sequel.

**Lemma 3.3** *Let $(X_n)_{n \geq 1}$ be a sequence of independent identicaly distributed random variables such that $n\mathbb{P}(|X_1| > n) = o(1)$ and $(T_n)_{n \geq 1}$ be a sequence of positive integer valued random variables satisfying $\frac{T_n}{n} \to c$ in probability, with $0 < c < \infty$. Then, denoting by $S_n = \sum_{i=1}^n X_i$, we have*

$$\lim_{n \to \infty} \left( \frac{S_{T_n}}{T_n} - \mathbb{E}(X_1 \mathbb{1}_{\{|X_1| \leq n\}}) \right) = 0 \quad \text{in probability.}$$

*Proof:* This is a standard extension of the strong law of large numbers and can be found, for instance, as theorem 5.2.6 of [3]. □

*Proof of theorem 1.1:* It is enough to study the $T \to \infty$ limit of all individual terms appearing in the expression for $f_T$ given in lemma 3.2.

Let $X_j = \log(1 + 2\beta V_j) \leq 0$. Because $\mathbb{P}(X_j > T)$ vanishes superexponentially fast, it follows that $T\mathbb{P}(|X_1| > T) = o(1)$ and we can apply lemma 3.3 to show that the first term converges in probability,

$$\frac{1}{2\beta T} \sum_{j=1}^{N_T} \log(1 + 2\beta V_j) \stackrel{\text{prob}}{\to} \frac{1}{2\beta} p \mathbb{E}[\log(1 + 2\beta V_1)].$$

The second term trivialy vanishes in the limit. As for the third term, remark that $1 \leq T - U_N < T$, therefore it also vanishes in the limit. For the fourth term, remark that $(T - U_N)^{-1} + 2\beta \leq 1 + 2\beta$ and $\sum_{j=1}^{N_T}(V_j^{-1} + 2\beta)^{-1} \leq \sum_{j=1}^{N_T} V_j = U_{N_T} < T$, hence this term gives also a vanishing limiting contribution. The fifht term yields, by definition, $z$.

The sixth term involves the study of the random variable $\Delta_T = T - U_N$. Obviously, $\mathbb{P}(\Delta_T \geq k) = \mathbb{P}(U_{N_T+k} - U_{N_T} = 0)$. Now, since $\Delta_T < T$, this random variable cannot be a plain geometrical one. In the limit $T \to \infty$, it can be shown however (see formula 6.5, page 194 of [7] for instance) that $\mathbb{P}(\Delta_\infty \geq k) = q^{k-1}$, for $k = 1, 2, \ldots$. Hence, the sixth term yields

$$\lim_{T \to \infty} \frac{1}{2\beta T} \frac{b^2}{T} \frac{U_{N_T}}{T - U_{N_T}} = \frac{z}{2\beta \Delta_\infty} \quad \text{in distribution.}$$

Because $(V_i^{-1} + 2\beta)^{-1} \geq (1 + 2\beta)^{-1}$, it follows that

$$\left( \sum_{i=1}^{N_T} ((V_i^{-1} + 2\beta)^{-1} \right)^{-1} \leq (1 + 2\beta) N_T^{-1}$$

and since $N_T \to \infty$ this inverse sum vanishes in the limit. Hence the seventh term converges, in distribution, to $-\frac{2\beta z \Delta_\infty}{1+2\beta z \Delta_\infty}$.

For the next term we use again the lemma 3.3 to show that

$$\frac{2b^2 \beta}{T} \frac{1}{T} \left( \sum_{i=1}^{N_T} (V_i^{-1} + 2\beta)^{-1} \right) \stackrel{\text{prob}}{\to} -2pz\mathbb{E}[(V_i^{-1} + 2\beta)^{-1}].$$

For the following term we obtain,

$$-2\frac{b^2}{T} \frac{[\frac{U_{N_T}}{T}(1+2\beta\Delta_T)^{-1} - \frac{1}{T}\sum_{i=1}^{N_T}(V_i^{-1}+2\beta)^{-1}]}{1+[(2\beta+\Delta_T^{-1})\sum_{i=1}^{N_T}(V_i^{-1}+2\beta)^{-1}]^{-1}}.$$

Again, use the observations that $2\beta + \Delta_T^{-1} \geq 2\beta + 1$ and $\sum_{i=1}^{N_T}(V_i^{-1}+2\beta)^{-1} \geq N_T(1+2\beta)^{-1}$ so that the big denominator $1 + [\cdots]$ reduces to 1 in the limit $T \to \infty$. Subsequently, this term yields:

$$-2z\left[(1+2\beta\Delta_\infty)^{-1} - p\mathbb{E}[(V_1^{-1}+2\beta)^{-1}]\right] \quad \text{in distribution.}$$

The same arguments can be used to study the last term

$$-\frac{b^2}{2\beta T(T-U_{N_T})^2} \left[2\beta + \Delta_T^{-1} + \left(\sum_{i=1}^{N_T}(V_i^{-1}+2\beta)^{-1}\right)^{-1}\right]^{-1}.$$

Obviously, again, the term $\left(\sum_{i=1}^{N_T}(V_i^{-1}+2\beta)^{-1}\right)^{-1}$ vanishes in the limit. It remains a contribution of the form

$$-\frac{z}{2\beta\Delta_\infty} + \frac{z}{1+2\beta\Delta_\infty}.$$

Adding together all the terms, the contributions of all but the first one conspire to disappear. $\square$

## 4 Finite volume ground state energy

We can now consider the behaviour of the ground state energy. First notice that a standard Laplace's argument guarantees that

$$-\lim_{\beta \to \infty} \frac{1}{\beta} \log \left[\mathbb{E}\left(\exp(-\beta H_T)\right)|(N,U))\right] = \inf_{X \in \mathcal{C}_{[0,T]}^{0,b}} H_T(U,X).$$

*Proof of the theorem 1.2:* the proof proceeds by examining how the individual terms in the expression of $F_T(\beta) = -\frac{1}{\beta}\log Z_T(\beta) = Tf_T(\beta)$, where $f_T(\beta)$ is given in lemma 3.2, behave in the limit $\beta \to \infty$. Namely, we examine the behaviour when $\beta \to \infty$ of the individual terms for the expression below:

$$F_T(\beta) = Tf_T(\beta) = \frac{1}{2\beta} \sum_{i=1}^{N_T} \log(1 + 2\beta V_i) + \frac{1}{2\beta} \log T - \frac{1}{2\beta} \log(T - U_N)$$

$$- \frac{1}{2\beta} \log \left[ 1 + \left((T - U_N)^{-1} + 2\beta\right) \sum_{i=1}^{N_T} (V_i^{-1} + 2\beta)^{-1} \right] + b^2$$

$$+ \frac{1}{2\beta} \frac{U_{N_T}}{(T - U_{N_T})} - \frac{A_1(\beta)}{\beta} - \frac{A_2(\beta)}{\beta} - \frac{A_3(\beta)}{\beta} - \frac{A_4(\beta)}{\beta}.$$

It is easy to see that the sole terms remaining in the limit $\beta \to \infty$ are the fifth one — giving trivialy a contribution equal to $b^2$ — and the seventh one — namely,

$$-\lim_{\beta \to \infty} \frac{A_1(\beta)}{\beta} = -\lim_{\beta \to \infty} \frac{b^2}{1 + (2\beta\Delta_T)^{-1} + \left[\sum_{i=1}^{N_T} \left(1 + (2\beta V_i)^{-1}\right)^{-1}\right]^{-1}} = -\frac{b^2}{1 + N_T^{-1}}.$$

Adding the two non vanishing contributions, we get

$$\lim_{\beta \to \infty} F_T(\beta) = b^2 - \frac{b^2}{1 + N_T^{-1}} = \frac{b^2}{1 + N_T}$$

and the theorem follows. □

*Proof of corollary 1.3:* This is obtained by an elementary summation:

$$\mathbb{E}\left(\frac{1}{N_T + 1}\right) = \frac{1}{pT} \sum_{n=0}^{T-1} \binom{T}{n+1} p^{n+1} q^{T-(n+1)} = \frac{1}{pT}(1 - q^T).$$

Therefore, collecting all terms,

$$\mathbb{E}\left(\inf_{X \in \mathcal{C}_{[0,T]}^{0,b}} H_T(U, X)\right) = \mathbb{E}\left(\lim_{\beta \to \infty} F_T(\beta)\right) = \frac{b^2}{pT}(1 - q^T).$$

□

# 5 Conclusion and open problems

It is established that the model studied here has well defined thermodynamics in the sense that the infinite volume limit of the quenched specific free energy exists; moreover, this limit is a trivial random variable — *i.e.* almost surely a constant. This result shows the sample independence of the thermodynamic properties of the model at every temperature; a feature common to all short range disordered systems.

The large scale behaviour of the average ground state energy is also very instructive. Suppose first that the parameters $p$ and $q$ are fixed — independent of $T$ — and non trivial *i.e.* $0 < p, q < 1$. Then,
$$\mathbb{E}[\inf_X H_T(U, X)] \sim \frac{z}{p} \ \text{ for large } T$$
and it has a finite value only when $z < +\infty$. Now, a finite and strictly positive $z$ means that the total charge scales as $T^{1/2}$ for large $T$, while the number of visited sites from the directed random walks behaves like $pT$. On the other hand there is no obstruction in choosing $p$ and $q$ depending on $T$. To illustrate what it may happen, suppose that we choose $p$ to scale asymptotically as $\alpha/T$, with $0 < \alpha < +\infty$. Then

$$\mathbb{E}[\inf_X H_T(U, X)] \sim \frac{b^2}{\alpha}(1 - \exp(-\alpha))$$

and this expectation is finite provided that the total charge *remains finite*. In the same time, the total number of visited sites does not diverge. Of course various other asymptotic behaviours can be obtained by choosing different scaling laws for the parameters $p$ and $q$. The result we have obtained allows an exhaustive study of all these scaling behaviours.

What remains an open problem for the moment is the statistical mechanics study of the system. Namely, for every realisation $U$ of the random environment, we can define a random finite volume Gibbs measure, $\mu_T$, on the space $\mathcal{C}^{0,b}_{[0,T]}$ as a measure absolutely continuous with respect to the standard Gaussian measure, $\gamma_T$, of the Brownian bridge. This Gibbs measure can be defined through its Radon-Nikodým derivative

$$\frac{d\mu_T}{d\gamma_T}(U, X) = \frac{\exp(-\beta H_T(U, X))}{Z_T(\beta)},$$

where $Z_T(\beta)$ is the partition function that depends also on $U$. It is not yet clear whether a Dobrushin-Lanford-Ruelle construction is possible for this model. Neither is it clear whether there is a phase transition or not, should it be in the DLR or the weak sense. These fundamental questions for the statistical mechanics formulation of the model are presently under investigation.

A second problem that is still under investigation concerns the thermodynamics of the model defined for genuine random walks and not directed random walks. This destroys the explicit renewal process structure of the model and makes the computations much more complicated but not untractable.

Summarising, inspired from the protein folding statistical mechanics problems, we are confronted with a new class of systems, reminiscent of both spin glasses and random walks in random environments.

# A   Identities for the Brownian bridge

Let $(W_s : s \geq 0)$ be a Brownian Motion starting from 0 and introduce the process $(Y_s : s \geq 0)$ by

$$Y_t = (T-t) \int_0^t \frac{dW_s}{T-s}.$$

Then, the Brownian bridge between $X_0 = 0$ and $X_T = b$ is expressed by

$$X_t = b\frac{t}{T} + Y_t \text{ for } t \in [0,T].$$

It is immediate to verify the following identities that are reported here only for completeness

$$\mathbb{E}(Y_t) = 0, \ \mathbb{E}(X_t) = \frac{bt}{T}, \ \mathbb{E}(X_t - X_s) = \frac{b}{T}(t-s)$$

and

$$\mathbb{E}(Y_t^2) = \frac{T-t}{T}t, \ \mathbb{E}(X_t^2) = \frac{b^2 t^2}{T^2} + \frac{(T-t)}{T}t.$$

For $s \leq t$, we have also

$$\mathbb{E}(Y_s Y_t) = \frac{(T-t)}{T}s, \ \mathbb{E}(X_s X_t) = \frac{b^2 st}{T^2} + \frac{(T-t)}{T}s,$$

$$\mathbb{E}((Y_t - Y_s)^2) = \frac{(t-s)}{T}(T-(t-s)), \ \mathbb{E}((X_t - X_s)^2) = \left(\frac{b^2}{T^2} - \frac{1}{T}\right)(t-s)^2 + (t-s) \leq t,$$

and

$$\operatorname{Var}((X_t - X_s)) = (t-s) - \frac{1}{T}(t-s)^2.$$

Finally, for $s_1 < s_2 \leq t_1 < t_2$, we have

$$\mathbb{E}((Y_{t_2} - Y_{t_1})(Y_{s_2} - Y_{s_1})) = -\frac{(s_2 - s_1)(t_2 - t_1)}{T},$$

$$\mathbb{E}((X_{t_2} - X_{t_1})(X_{s_2} - Y_{s_1})) = \left(\frac{b^2}{T^2} - \frac{1}{T}\right)(s_2 - s_1)(t_2 - t_1),$$

and

$$\operatorname{Cov}\left((X_{t_2} - X_{t_1}), (X_{s_2} - X_{s_1})\right) = -\frac{1}{T}(s_2 - s_1)(t_2 - t_1).$$

# B   Determinant and positive definiteness for a class of matrices

For $\vec{\ell} = (\ell_1, \cdots, \ell_n) \in \mathbb{R}^n$ denote by $D(\vec{\ell})$ the diagonal matrix, $D(\vec{\ell})_{ij} = \ell_i \delta_{ij}$ for $i,j = 1, \cdots, n$. For $a \in \mathbb{R}$ denote by $\tilde{a}$ the constant matrix whose elements are $\tilde{a}_{ij} = a$ for

$i, j = 1, \cdots, n$. From [13], page 92, we get the following property,

$$\det (D(\vec{\ell}) + \tilde{a}) = \prod_{j=1}^{n} \ell_j + a \sum_{i=1}^{n} \prod_{\substack{j=1 \\ j \neq i}}^{n} \ell_j.$$

Consequently, if $\ell_j \neq 0$ for every $j$, then

$$\det (D(\vec{\ell}) + \tilde{a}) = \left( \prod_{j=1}^{n} \ell_j \right) \left( 1 + a \sum_{i=1}^{n} \ell_i^{-1} \right). \tag{16}$$

Remark that the matrix $(D(\vec{\ell}) + \tilde{a})$ as well as all its minorants

$$(D(\vec{\ell}) + \tilde{a})^{(k)} = \begin{pmatrix} \ell_1 & & 0 \\ & \ddots & \\ 0 & & \ell_k \end{pmatrix} + \begin{pmatrix} a & \cdots & a \\ \vdots & & \vdots \\ a & \cdots & a \end{pmatrix}, \tag{17}$$

for $k = 1, \cdots n$ are symmetric and have the same form.

**Lemma B.1** *Assume that $\ell_i > 0$, for all $i = 1, \cdots, n$ and $a \geq 0$. Then,*

1. *all minorants satisfy $\det (D(\vec{\ell}) + \tilde{a})^{(k)} > 0$ for $k = 1, \cdots n$,*

2. *the matrix $(D(\vec{\ell}) + \tilde{a})$ is positive definite,*

3. *there exists a positive definite symmetric matrix $\Gamma(\ell, a)$ that satisfies*

$$(D(\vec{\ell}) + \tilde{a}) = \left( \Gamma(\vec{\ell}, a) \right)^2.$$

*Proof:* The two first assertions of the lemma follow immediately from relations (16) and (17). The third assertion follows from the positive definitness of the matrix $(D(\vec{\ell}) + \tilde{a})$. $\square$

**Corollary B.2** *Under the same conditions as for the previous lemma,*

$$\det \Gamma(\vec{\ell}, a) = \left( \prod_{j=1}^{n} \ell_j \right)^{\frac{1}{2}} \left( 1 + a \sum_{j=1}^{n} \ell_j^{-1} \right)^{\frac{1}{2}}.$$

# C   Inversion for a class of matrices

We use the notation already introduced in Appendix B. For $\vec{\ell} = (\ell_1, \cdots, \ell_n) \in \mathbb{R}^n$ we define the matrix $S(\vec{\ell})$ by $S(\vec{\ell})_{ij} = \ell_i \ell_j$ for $i, j = 1, \cdots, n$. Also we denote $\vec{\ell}^{-1} = (\ell_1^{-1}, \cdots, \ell_n^{-1})$, so that $D(\vec{\ell}^{-1}) = (D(\vec{\ell}))^{-1}$.

**Lemma C.1** *Assume $\ell_i \neq 0$ for every $i = 1, \cdots, n$.*

- *If $\gamma \in \mathbb{R}$ verifies $\gamma \sum_{i=1}^{n} \ell_i \neq 1$, then the matrix $(D(\vec{\ell}) - \gamma S(\vec{\ell}))^{-1}$ exists and verifies,*

$$(D(\vec{\ell}) - \gamma S(\vec{\ell}))^{-1} = D(\vec{\ell})^{-1} + \gamma \left(1 - \gamma \sum_{i=1}^{n} \ell_i\right)^{-1} \tilde{1}.$$

- *Reciprocally, if $\gamma \sum_{i=1}^{n} \ell_i^{-1} \neq -1$, then the matrix $(D(\vec{\ell}) + \gamma \tilde{1})^{-1}$ exists ans verifies,*

$$(D(\vec{\ell}) + \gamma \tilde{1})^{-1} = D(\vec{\ell}^{-1}) - \gamma \left(1 + \gamma \sum_{i=1}^{n} \ell_i^{-1}\right)^{-1} S(\vec{\ell}^{-1}).$$

*Proof:* Denote by $R(\vec{\ell})$ the matrix $R(\vec{\ell})_{ij} = \ell_j$, for $i, j = 1, ..., n$. It is immediate to verify the identities $D(\vec{\ell})R(\vec{\ell}) = S(\vec{\ell})$, $R(\vec{\ell})^2 = \left(\sum_{i=1}^{n} \ell_i\right) R(\vec{\ell})$, and $R(\vec{\ell})D(\vec{\ell})^{-1} = \tilde{1}$. Hence, $D(\vec{\ell}) - \gamma S(\vec{\ell}) = D(\vec{\ell})(I - \gamma R(\vec{\ell}))$ and therefore, it follows by direct computation $(I - \gamma R(\vec{\ell}))^{-1} = I + \gamma \left(1 - \gamma \sum_{i=1}^{n} \ell_i\right)^{-1} R(\vec{\ell})$. Now

$$D(\vec{\ell}) - \gamma S(\vec{\ell}))^{-1} = D(\vec{\ell})^{-1} + \gamma \left(1 - \gamma \sum_{i=1}^{n} \ell_i\right) R(\vec{\ell})D(\vec{\ell})^{-1}$$

which is enough to conclude. □

Hence, under the condition of this lemma and if $\gamma \neq 0$,

$$(D(\vec{\ell}) - \gamma S(\vec{\ell}))^{-1} = D(\vec{\ell}^{-1}) + \left(\gamma^{-1} - \sum_{i=1}^{n} \ell_i\right)^{-1} \tilde{1},$$

and

$$(D(\vec{\ell}) + \gamma \tilde{1})^{-1} = D(\vec{\ell}^{-1}) - \left(\gamma^{-1} + \sum_{i=1}^{n} \ell_i^{-1}\right)^{-1} S(\vec{\ell}^{-1}).$$

**Acknowledgments:** SM thanks the Insitut de Recherche Mathématique de l'Université de Rennes I and DP the Departamento de Ingeniería Matemática of the Universidad de Chile, where this paper was done. This work was partially supported by FONDECYT grant 1940405 and EU grants CI1-CT92-0046 and CHRX-CT93-0411.